\documentclass[aps,twocolumn,showpacs,preprintnumbers,amsmath,amssymb,superscriptaddress]{revtex4}
\usepackage{graphicx}
\usepackage{dcolumn}
\usepackage{bm}
\graphicspath{{./Figures/}}

\begin{document}

\preprint{APS/123-QED}

\title{Liquid phase stabilization versus bubble formation at a nanoscale-curved interface}
\author{Jarrod Schiffbauer}
\email{jschiffb@nd.edu}
\affiliation{University of Notre Dame, Department of Aerospace and Mechanical Engineering, Notre Dame, IN, 46556}
\author{Tengfei Luo}
\email{tluo@nd.edu}
\affiliation{University of Notre Dame, Department of Aerospace and Mechanical Engineering, Notre Dame, IN, 46556}
\affiliation{Center for Sustainable Energy at Notre Dame, Notre Dame, IN, 46556}

\date{\today}

\begin{abstract}
We investigate the nature of vapor bubble nucleation near a nanoscale-curved convex liquid-solid interface using two models; an equilibrium Gibbs model for homogenous formation, and a non-equilibrium dynamic van der Waals/diffuse interface model for phase change in an initially cool liquid. Vapor bubble formation is shown to occur for sufficiently large radius of curvature and is suppressed for smaller radii. Solid-fluid interactions are accounted for and it is shown that liquid-vapor interfacial energy--hence Laplace pressure--has limited influence over bubble formation. The dominant factor is the energetic cost of creating the solid-vapor interface from the existing solid-liquid interface, as demonstrated via both equilibrium and non-equilibrium arguments.
\end{abstract}

\maketitle
\begin{figure}
  \includegraphics[width=3.0 in]{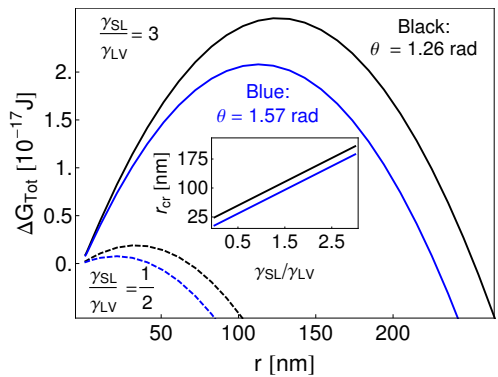}\\
  \caption{The Gibbs free energy change, Eqn.~\ref{eqn:e1}, at $T=0.85T_c$ for Argon on a gold-like nanosphere for $\tilde{\delta r}=1$ nm, for hydrophilic ($\theta=1.26$ rad) and hydrophobic ($\theta=1.57$ rad) surfaces with different strengths of solid-fluid interaction, as determined by $\gamma_{SL}/\gamma_{LV}$. (Inset) The critical radius estimates for the same angles as a function of $\gamma_{SL}/\gamma_{LV}$.}\label{fig:f2}
\end{figure}
\begin{figure}
  \includegraphics[width=3.35 in]{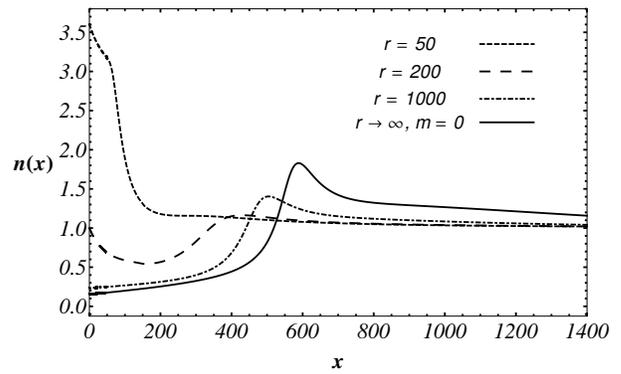}\\
  \caption{Density profile at $t=3500$ for several different inner radii $r$ showing liquid phase stabilization at small radius and bubble formation and expansion at large radius, converging to the planar case as $r\rightarrow\infty$. The viscosity is $\nu=1.67\times 10^{-6}$ m$^2$/s and $\alpha=0.208$, $\beta=47.192$, $\delta=10^{-3}$, and $\chi=-10^{-3}$, a weakly hydrophilic surface.}\label{fig:f3}
\end{figure}
\begin{figure}
  \includegraphics[width=3.35 in]{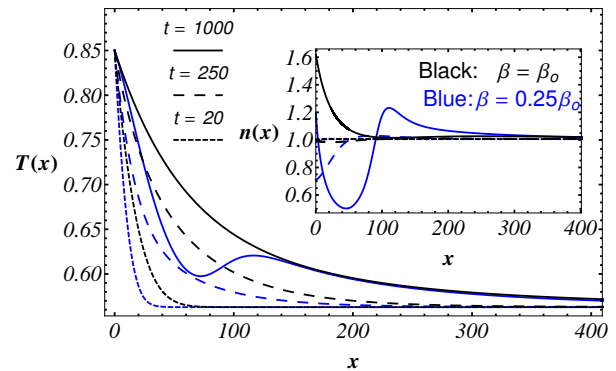}\\
  \caption{Several snapshots of the temperature profiles for $\beta_o=47.192$ and $0.25\beta_o$, and $\chi=-10^{-3}$. Inset shows a comparison of the density profiles near the interface for $t=1000$. Calculations for a range of $\delta=10^{-9}-10^{-1}$ show essentially the same behavior.}\label{fig:f4}
\end{figure}
\begin{figure}
  \includegraphics[width=3.35 in]{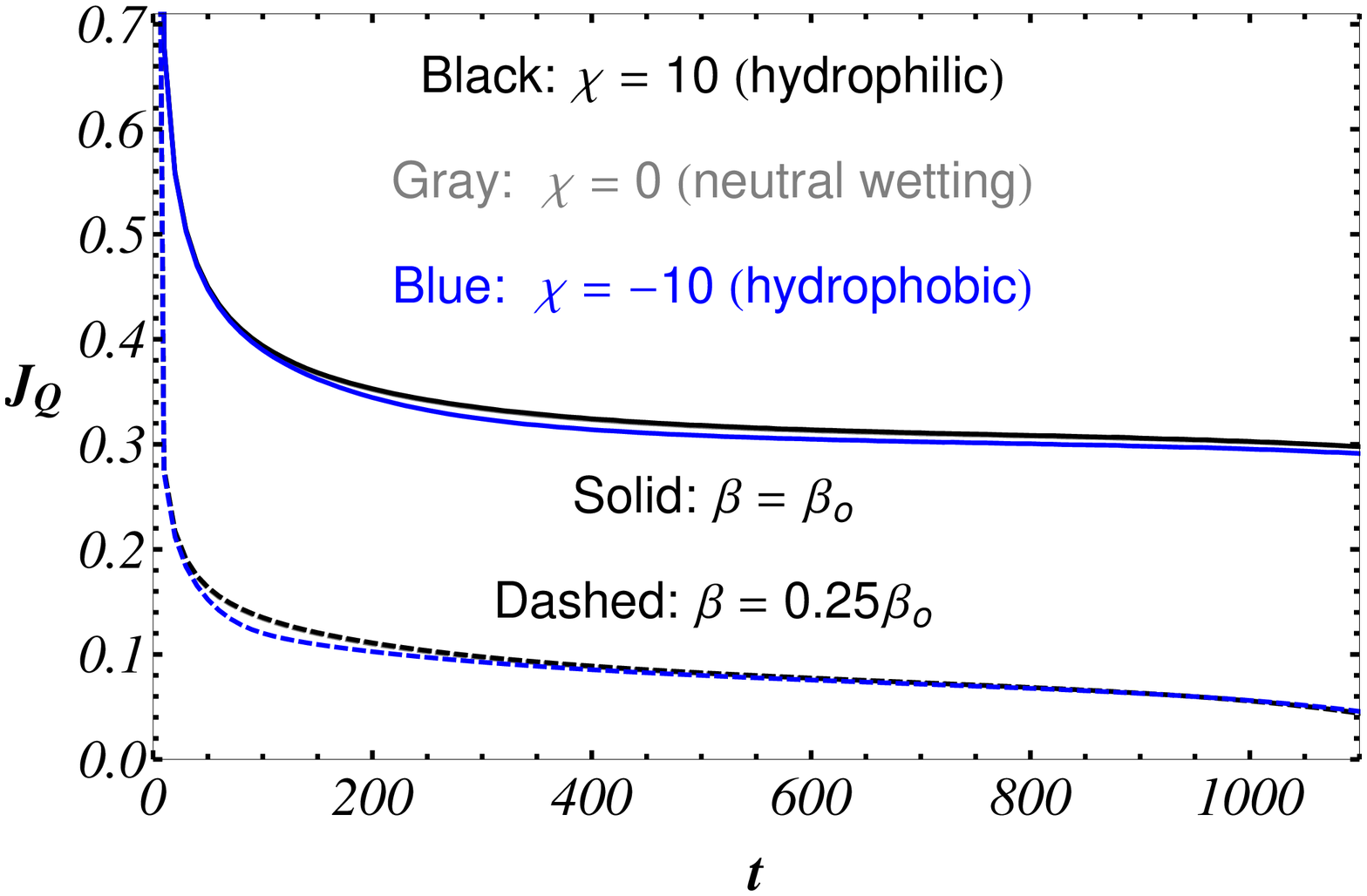}\\
  \includegraphics[width=3.25 in]{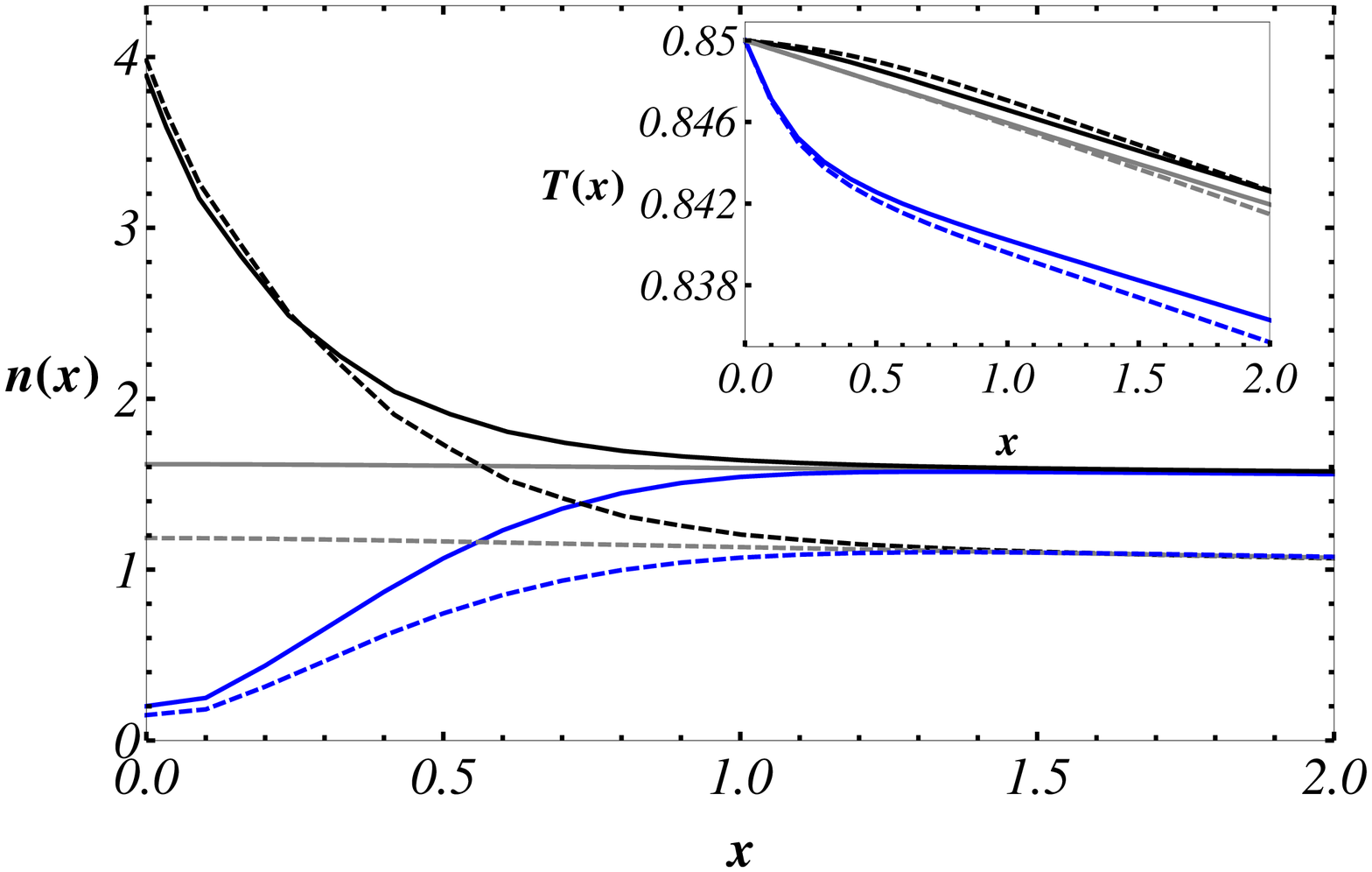}\\
  \caption{Heat flux (top) and interfacial density profiles (bottom) for strongly hydrophilic, $\chi=10$, strongly hydrophobic, $\chi=-10$, and neutral, $\chi=0$ interfaces. Inset shows a comparison of the density profiles near the interface for $t=1000$. Calculations are shown for $\delta=10^{-2}$. The surface interaction enters in the capillary terms and is much weaker for small $\delta$.}\label{fig:f5}
\end{figure}

A number of potentially transformative technologies rely on energy and momentum transfer between a hot, nanostructured solid and a surrounding fluid~\cite{Hartland2002,Chen2013}. These span applications as diverse as plasmonic photothermal cancer therapy~\cite{ElSayed2008,Lapotko2011,Drezek2011,Bischof2012}, photocatalysis~\cite{Shetty2014,Ye2016}, solar-powered water desalination~\cite{Polman2013,Halas2013,Tiwari2016,Zhu2016}, and physico-chemical separations~\cite{Halas2015}. Metallic~\cite{Braun2002,Hartland2004}, metal-dielectric structured~\cite{LizMarzan2003}, and/or molecularly functionalized nanoparticles (NPs)~\cite{Pradeep2003,Gazelter2016} have been studied as candidates for many of these applications. Nanoscale cavitation-oscillation~\cite{Kotaidis2005,Lapotko2010, Sasikumar2014, Hou2014, LombardPRL} and phase-change phenomena~\cite{Merabia2009} have been reported, but a simple thermodynamic criterion for vapor bubble formation, analogous to nanocrystal nucleation in a melt~\cite{Kaptay2011} (or cavitation from a \emph{bulk} fluid), is conspicuously lacking. While it is known that interfacial forces and molecular structure affects heat transport~\cite{TZhang2016, RamosAlvarado2016}, the roles of interfacial energies, viscous dissipation, and phase change near nanoscale interfaces~\cite{EvansPRL,Binder1995, Merabia2009,Keblinski2014} remain somewhat unclear.\\
\indent Nanoscale-confined phase stabilization due to surface-fluid interaction has been predicted under a variety of circumstances (see ~\cite{EvansPRL,Binder1995} for example). However, the phase stability of a liquid layer surrounding small gold nanoparticles observed in molecular dynamics simulations~\cite{Merabia2009,Keblinski2014} is quite curious; it is not \emph{a priori} clear how the curvature of a convex solid-fluid interface would stabilize the heated fluid against vaporization. One hypothesis is that the Laplace pressure required to sustain a stable bubble of small radius is too high and this suppresses vapor formation at the interface~\cite{Merabia2009}. Here we examine this by considering both equilibrium vapor formation criteria and non-equilibrium hydrodynamic calculations. However, we frame the discussion in terms of interfacial energies, rather than Laplace pressure~\cite{Kaptay2011}.\\
\begin{figure}
  \includegraphics[width=2.25 in]{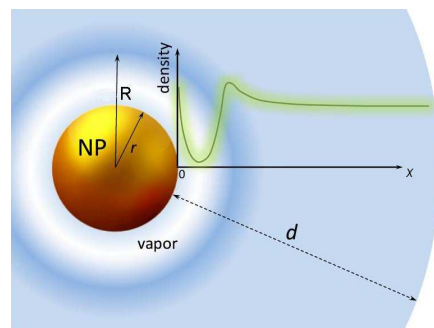}\\
  \caption{Schematic of coordinates for system showing inner surface at $r$, vapor bubble between $r$ and $R$, and the bulk-like liquid. Note $d$ is chosen to be large enough that any pressure waves reflected from the outer boundary, where the pressure is held constant, do not have sufficient time to return to the bubble region during the course of simulation.}\label{fig:f1}
\end{figure}
\indent As a first model, consider the equilibrium formation of a vapor layer of thickness $\tilde{\delta r}$\footnote{A `~' is used to denote dimensional variables. Non-dimensional variables, as well as constants, are left unadorned.} from a liquid at uniform temperature $\tilde{T}$ surrounding a particle of radius $\tilde{r}$ (see Fig.~\ref{fig:f1}). The change in Gibbs free energy is the (reversible) free energy change associated with an input heat of $\tilde{T}\Delta\tilde{S}_{LV}$ turning liquid near an existing solid-liquid interface into a vapor layer with a solid-vapor and a liquid-vapor interface,
\begin{eqnarray}& \Delta \tilde{G}_{total} =  \frac{4\pi}{3}\frac{(\Delta \tilde{H}_{LV}-\tilde{T}\Delta \tilde{S}_{LV})(\tilde{R}^3-\tilde{r}^3)}{\tilde{V}_V} \nonumber \\ & + 4\pi\left[\frac{\gamma_{SV}(\tilde{R}^3-\tilde{r}^3)}{\tilde{r}}+\frac{\gamma_{LV}(\tilde{R}^3-\tilde{r}^3)}{\tilde{R}}\frac{\tilde{V}_L}{\tilde{V}_V}-\frac{\gamma_{SL}(\tilde{R}^3-\tilde{r}^3)}{\tilde{r}}\frac{\tilde{V}_L}{\tilde{V}_V}\right]\label{eqn:e1}\end{eqnarray}
Here, $\tilde{R}=\tilde{r} +\tilde{\delta r}$, and the phase (solid, liquid, vapor) for the molar volumes, $\tilde{V}$, changes in enthalpy $\tilde{H}$ and entropy $\tilde{S}$, as well as interfacial tensions $\gamma$, are denoted by respective subscripts $S$, $L$, and $V$. Minimizing Eqn.~\ref{eqn:e1} with respect to radius, $\tilde{r}$, keeping terms leading order in $\tilde{\delta r}$, and using Young's equation, $\gamma_{SV}=\gamma_{LV}\cos\theta + \gamma_{SL}$,  yields a critical particle radius for formation of a thin vapor layer ($\tilde{\delta r}<<\tilde{r}$),
\begin{equation}\tilde{r}_{cr}=\frac{\gamma_{LV}\left(5\tilde{V}_V\cos\theta+3\tilde{V}_L\right)+5\gamma_{SL}\Delta \tilde{V}_{LV}}{-2(\tilde{T}\Delta \tilde{S}_{LV} -\Delta \tilde{H}_{LV})} \end{equation}
Consider Eqn.~\ref{eqn:e1} with $\gamma_{LV}=3.4$ mN/m (Argon) for a range of contact angles (with respect to the flat, equilibrium interface) and ratios of $\gamma_{SL}/\gamma_{LV}$ (see Fig.~\ref{fig:f2}). Increasing the solid-fluid surface energy corresponds to a higher barrier to vapor formation for radii below $\Delta \tilde{G}_{total} = 0$ and greater release of energy above. Similarly, hydrophilic surfaces, $\theta<\pi/2$, have a higher barrier to vapor-layer formation than hydrophobic surfaces, $\theta>\pi/2$, and result in a smaller release of energy for spontaneous transition.\\
\indent Comparing the mercury-liquid interfaces for several liquids with various liquid-vapor interfaces~\cite{AdamsonGast}, the solid-vapor and solid-liquid interfacial energies for fluid-metal interfaces can be estimated to be about an order of magnitude larger than typical liquid-vapor interfacial energies. Below the critical temperature, $T_c$, the ratio of molar volumes will be $<1$; for example, Argon has ratios ranging from $10^{-3}-10^{-1}$ up to about $T=0.85T_c$ ($128$ K)~\cite{NISTsite}. So the solid-liquid and liquid-vapor contributions to Eqn. 1 are generally smaller than the solid-vapor contribution. The critical radius (inset, Fig.~\ref{fig:f2}) increases with increasing solid-vapor interfacial energy. Even in the limit of vanishing liquid-vapor interfacial tension, a non-zero solid-vapor interfacial energy can inhibit vapor formation. Thus the dominant factor in vapor formation is not the liquid-vapor interfacial energy, but rather energetic cost of creating the solid-vapor interface.\\
\indent The energetic criterion for reversible formation of a vapor layer in a uniformly heated liquid is given by Eqn.~\ref{eqn:e1}. However, an initially cold liquid in contact with a hot particle will undergo heating, expansion, and phase change. Viscous dissipation has been shown to play an important role in non-equilibrium bubble dynamics~\cite{LombardPRE}. So the Gibbs approach can provide a lower bounds for the energy required to nucleate a bubble. But a non-equilibrium theory is required to shed light on the dynamics of vaporization in an initially cool liquid placed in contact with a hot solid. Because the diffuse interface/dynamic van der Waals theory~\cite{HohenbergRMP,TurskiPRA, RowlWidom, AndersonNIST,BabinJCP,BabinJPC,OnukiPRE, TeshegawaraPRE} has been used previously to study bubble dynamics~\cite{LombardPRL,LombardPRE}, in good agreement with the classical Rayleigh-Plesset equation, MD simulations~\cite{Sasikumar2014}, and experiments~\cite{Kotaidis2005} of bubble growth and collapse, we adopt a variation on this model. We wish to isolate the roles of viscous dissipation, capillary forces, and interfacial curvature on the phase change and heat transfer properties of the fluid from considerations of energy transport and capacity of the solid. Therefore we consider the evolution of an initially uniform fluid held between rigid, impenetrable surfaces of infinite interfacial conductance (see Fig.~\ref{fig:f1}) in thermal equilibrium with infinite capacity baths at fixed temperatures $T(r)=0.85T_c$ and $T(r+d)=0.56T_c$.\\
\indent The model is formulated using hydrodynamic conservation equations for mass, momentum, and energy~\cite{BabinJCP,BabinJPC,OnukiPRE} and boundary conditions which are solved numerically (see Supplemental Materials~\cite{Supplement}). The one-density van der Waals theory includes a density gradient contribution to the free energy density~\cite{RowlWidom, OnukiPRE, TeshegawaraPRE}. The elements of the pressure tensor are then $\tilde{P}_{i,j}=\left[\tilde{n}k_B\tilde{T}/(1-\Omega_o\tilde{n})-\varepsilon\Omega_o\tilde{n}^2-C\tilde{T}\tilde{n}\nabla^2\tilde{n}+C\tilde{T}\left(\nabla\tilde{n}\right)^2\right]\hat{\delta}_{i,j}+C\tilde{T}\partial_i\tilde{n}\partial_j\tilde{n}$ where $\Omega_o$ and $\varepsilon$ are respectively the Lennard-Jones volume and well-depth, $k_B$ is the Boltzmann constant, and $C$ is the density gradient coupling parameter, which we discuss subsequently. We neglect angular non-uniformity in surface temperature and restrict ourselves to cases with radial symmetry. The divergence operator is $\mathbf{\mathcal{D}}_m\left[\varphi(x)\right]= x^{-m}\partial_x\left[x^m \varphi(x)\right]$ with $m=0$ representing planar, and $m=2$ spherical, symmetry. The (dimensionless) 1D governing equations are: $\partial_tn+\mathbf{\mathcal{D}}_m\left[nv\right]=0$ and $n\partial_tv+nv\partial_xv=\mathbf{\mathcal{D}}_m\left[D_{xx}-P_{xx}\right]$ where we have defined\\ \begin{eqnarray} D_{xx}&-&P_{xx}=\frac{9}{8}\alpha n^2\nonumber \\-\frac{1}{3}\frac{nT}{1-\alpha n}&+&\delta nT\left(\frac{m\partial_xn}{x}+\partial^2_{xx}n\right) \label{eqn:e3}\\-\frac{3}{2}\delta T\left(\partial_xn\right)^2  &+&\frac{7}{3}\beta n\partial_xv + \frac{1}{3}\frac{m\beta nv}{x} \nonumber \end{eqnarray} and the temperature profile is given by \begin{eqnarray}\frac{3}{2}n\partial_tT&=&\mathbf{\mathcal{D}}_m\left[\beta n\partial_xT\right]-\frac{nT}{1-\alpha n}\mathbf{\mathcal{D}}_m\left[v\right]-\frac{3}{2}nv\partial_xT\nonumber\\ &+& \frac{7}{9}\beta n\left(\partial_xv\right)^2+\frac{1}{9}\frac{m\beta nv\partial_xv}{x}\label{eqn:e4}\end{eqnarray}
\indent The Lennard-Jones parameters for Argon $\varepsilon = 7.033\times 10^{-21}$ J and $\Omega_o^{1/3}=0.345$ nm set the energy and length scales. The number density is scaled by the liquid bulk number density, $n_{LB}$ at pressure, $100$ kPa, and temperature $T=0.56T_c$, where $T_c$ normalizes the temperature. Velocity and time scales may then be defined, $v_o=\sqrt{3k_BT_c/M}=97$ m/s and $\tau_o=\Omega_o^{1/3}/v_o= 3.5$ ps. Assuming a constant kinematic viscosity, $\nu$, three dimensionless parameters describe the dynamics: the excluded volume fraction $\alpha=\Omega_on_{LB}$, an inverse Reynold's number $\beta=\nu/\Omega_o^{1/3}v_o$ controlling viscous dissipation and thermal conductivity, and a capillarity parameter $\delta=Cn_{LB}/3k_B\Omega_o^{2/3}$.\\
\indent The gradient coupling parameter, $C$, is related to the equilibrium liquid-vapor surface tension through the action integral over the density~\cite{RowlWidom,deGennes}, assuming $C$ is a constant~\cite{TeshegawaraPRE,LombardPRL,LombardPRE}. Note, however, for typical substances~\cite{NISTsite}, the values of $C$ thus obtained vary by over an order of magnitude with temperature. While $n$, $P$, and $T$ are held constant at the exterior (cold) boundary, the interior boundary condition, $\partial_xn|_{x=r}=-\chi/T(r)$ accounts for adsorption at the solid surface where $\chi=\phi_o\Omega_o^{1/3}/CT_cn_{LB}$ with wetting potential $\phi_o$~\cite{TeshegawaraPRE,LombardPRE}. $\phi_o$ can be related to the contact angle for the appropriate liquid-vapor-solid equilibrium for a flat interface~\cite{LombardPRE}, again assuming independence of state. For simplicity, we treat $\beta$, $\delta$, and $\chi$ as free parameters with a range informed by data for Argon~\cite{NISTsite}. A more physically realistic model would include a state-dependent viscosity (see for instance~\cite{LombardPRL}). But for a constant capillary coupling, across the range of physically reasonable values for viscosity~\cite{NISTsite}, bulk-like energy and momentum transport will dominate the leading-order dynamics over capillary forces.\\
\indent Numerical solutions were obtained for a range of parameters, $\delta \sim 10^{-9}$ to $10^{-1}$, $\|\chi\|\sim-10^{-3}$ to $10$, and $\beta\sim 10$ to $10^2$. Here we summarize and discuss the main results and compare them to the Gibbs model. Density profiles $n(x,t)$ are shown in Fig.~\ref{fig:f3} for several different inner radii $r$ at $t=3500$ for $\delta=10^{-3}$ and weakly hydrophilic surface, $\chi=10^{-3}$. The density for $r=50$ (about $\tilde{r}= 18$ nm) does not exhibit vapor formation. Instead, a layer of high-density liquid forms adjacent to the heated surface. For larger radii (between $150\leq r\leq200$), a liquid-vapor interface begins to develop. As $r$ increases, the spherical $m=2$ model approaches planar $m=0$ behavior. The model predictions are within a reasonable range of Gibbs radius for Argon with weak surface-liquid forces (see Fig.~\ref{fig:f2} inset.)\\
\indent Since $\delta\sim O(\gamma_{LV}^2)$, the range $\delta=10^{-9}$ - $10^{-1}$ corresponds to a range of 4 orders of magnitude in equilibrium liquid-vapor energy. Physically reasonable values for Argon are at the upper end of this range. Changing $\delta$ alone has no appreciable effect on vapor layer formation. However, modest change in $\beta$ has a pronounced effect (Fig.~\ref{fig:f4}). This is unsurprising. The capillary terms in Eqn.~\ref{eqn:e3} give higher-order contributions to the dynamics. In the present model, non-equilibrium phase stabilization originates in the $x^{-1}$ radial contribution to the viscous dissipation, represented by the last terms in each of Eqns.~\ref{eqn:e3} and~\ref{eqn:e4}. In the limit $r\rightarrow\infty$, the radial component of the dissipation vanishes and a vapor layer forms regardless of $\beta$.\\
\indent For larger $\beta$/smaller $r$, heat conduction near the interface is rapid, resulting in a broader temperature distribution and dissipative losses are high (Fig.~\ref{fig:f4}). According to the Gibbs formulation in Eqn.~\ref{eqn:e1}, the magnitude of the free energy change required for layer formation at a sub-equilibrium radius increases with increasing thickness, $\tilde{\delta r}$, of the hot layer. This suggests, despite the local temperature, the fluid close to the interface cannot accumulate enough internal energy to undergo phase change. The liquid-vapor interfacial tension, via realistic values of the capillary parameter, however, appears to have little effect on the dynamics of the model; whereas the viscosity, which gives rise to dissipative dynamics, obviously does not appear in the Gibbs formulation. Nonetheless, the picture which emerges from the non-equilibrium calculations is consistent with the equilibrium criterion in as much as the dissipative terms compete with the creation of the solid-vapor interface.\\
\indent Following Lombard et. al.~\cite{LombardPRE}, the solid-fluid forces are parameterized by the contact angle, yielding a wetting potential $\phi_o$~\cite{TeshegawaraPRE}. The model does not predict a phase stabilization effect based on the variation of $\chi$ alone, and vapor-layer formation was affected primarily by $\beta$ and $r$. The wettability, parameterized by $\chi$, influences interfacial heat transfer via the establishment of short-range density gradients (see Fig.~\ref{fig:f5}), increasing or decreasing local conductivity. Hydrophilic surface interactions show enhancement over neutral interfaces, while hydrophobic interfaces result in reduced heat transport, in qualitative agreement with analytic~\cite{Caplan2014} and molecular-dynamics~\cite{RamosAlvarado2016,Tascini2017} results.
There is some question regarding the ability of a continuum model to capture the physics of a fully-developed liquid-vapor interface of thickness approaching molecular dimensions~\cite{HoylstSOFT}. But this is not critical for initial vapor formation and early dynamics. The gradients $\partial_x n$ are initially small and the dissipation terms are dominant over capillary forces, since physically realistic $\delta<\beta$ by roughly 2 orders of magnitude across the temperature range~\cite{NISTsite}. Regarding surface forces, despite qualitative agreement with other results, the present treatment seems inadequate. Estimating the range of surface forces by the extent of hydrophobic and -philic boundary layers, there is no influence beyond about the L-J minimum, $2^{1/6}\Omega_o^{1/3}$. This is consistent with the repulsive component of a typical surface force, but the range of attractive forces is larger. Moreover, a continuous gradient cannot exist on a molecular length scale, except as an average over fluctuations. Lastly, the relationship between $\theta$ and $\phi_o$~\cite{LombardPRE} is based on the 3-phase equilibrium~\cite{RowlWidom,deGennes} but includes only 2 of the 4 equilibrium parameters appearing in Young's equation.\\
\indent The liquid density near the interface in the $r=50$ case (see Fig.~\ref{fig:f3}) is high compared to atomistic simulations~\cite{Merabia2009,Keblinski2014} and the apparent minimum radius about an order of magnitude larger than determined by MD~\cite{Keblinski2014}. But the two pictures agree qualitatively; there is a particle radius below which a superheated liquid layer remains at the interface and the liquid-vapor interface doesn't form. The discrepancies are attributed to surface-fluid interactions, as well as variation in viscosity with temperature. Interestingly, a number of experimental~\cite{Geim2016} and first-principles simulations~\cite{Grest2008,Bocquet2016} show modulation of the effective solvent viscosity and mass transport near nanoscale solid-fluid interfaces. While the correspondence between the solid-fluid and the liquid-vapor interfacial contribution to the Gibbs energy is clear, the non-equilibrium model makes no distinction, except through viscosity and density. Future work will aim to study effects of temperature-dependent viscosity, as well as surface-fluid forces via the inclusion of a body-force term in the momentum equation, in comparison to molecular simulations, in an effort to clarify this.\\
\indent In conclusion, we have demonstrated that liquid-vapor interfacial energy, hence Laplace pressure, does not suppress bubble formation. Both equilibrium and non-equilibrium arguments demonstrate the critical factor is instead the cost of creating the solid-vapor interface. Furthermore, it was demonstrated that rapid heat conduction and viscous dissipation near the curved interface can prevent bubble formation. The relationship between viscous dissipation and interfacial forces is of fundamental and practical interest for nanoscale systems, where fluctuations and molecular structure play important roles near interfaces, yet a continuum theory still captures the essential dynamics of the overall system.\\
\indent The results suggest that one can optimize nanoscale solid-fluid heat transfer and control nanoscale boiling by tailoring both geometry and surface properties, which affect the structure, hence viscosity, of adjacent fluid layers~\cite{RamosAlvarado2016}. The Gibbs criterion predictions regarding heat and momentum transfer at hydrophobic vs. hydrophilic nano-curved surfaces can be tested. This would not only be of fundamental interest, but regarding applications such as cancer PPT, solar desalination, or separations, controlling particle interfacial thermal and momentum transport through a combination of particle size and surface properties is of great practical value.\\

\begin{acknowledgments}
We gratefully acknowledge ND Energy for support through the ND Energy Postdoctoral Fellowship program and the Army Research Office, Grant No. W911NF-16-1-0267, managed by Dr. Chakrapani Venanasi\\
\end{acknowledgments}
\bibliography{bubbles}
%
%

%

\end{document}